\documentclass[aps,twocolumn,nofootinbib,superscriptaddress,preprintnumbers,pra,10pt]{revtex4-1}

\usepackage{float}
\usepackage{arydshln}
\usepackage{colortbl}
\usepackage{amsmath,amssymb}
\usepackage{dsfont} 
\usepackage{hyperref}
\usepackage{graphicx}
\usepackage{enumitem}
\usepackage{arydshln}
\usepackage{mathtools}
\usepackage{bbold}
\usepackage{soul}
\usepackage{slashed}
\usepackage{amsmath,bm}
\usepackage{yfonts}
\usepackage{lipsum}
\usepackage{slashed}
\usepackage{mathtools}
\usepackage{physics}
\usepackage{tikz}
\usepackage{tikz-feynman}

\makeatletter
\g@addto@macro\bfseries{\boldmath}
\makeatother

\newcommand{\be} {\begin{equation}}
\newcommand{\ee} {\end{equation}}
\newcommand{\bea} {\begin{eqnarray}}
\newcommand{\eea} {\end{eqnarray}}

\newcommand{\llpair}{\bar\ell\ell}

\allowdisplaybreaks

\begin{document}

\preprint{ZU-TH 50/25}
\title{Charm rescattering in $B^0\to K^0\llpair$: an improved analysis.}
 
\author{Gino Isidori }
\author{Zachary Polonsky}
\author{Arianna Tinari}

\affiliation{Physik-Institut, Universit\"at Z\"urich, CH-8057 Z\"urich, Switzerland}

\begin{abstract}
\vspace{5mm}
We improve upon previous explicit estimates of charm rescattering contributions to the decay $B^0\to K^0\llpair$ by including contributions from dipole interactions with the intermediate charm-meson states, and by further investigating the structure of the electromagnetic form factors. Using a model of fundamental meson fields inspired by heavy-hadron chiral perturbation theory, augmented by form factors motivated by theoretical considerations as well as experimental data, we provide a thorough investigation of rescattering contributions induced by intermediate $D^{(*)}D^{(*)}_s$ states.
\vspace{3mm}
\end{abstract}

\maketitle
\allowdisplaybreaks

\section{Introduction}

Flavor-changing neutral current (FCNC) transitions in the standard model (SM) provide excellent candidates for the observation of indirect evidence of beyond standard model (BSM) physics due to their heavy suppression in the SM. Of particular interest over the past decade has been the $b\to s\llpair$ transition: several analyses of exclusive-mode decays indicate significant tensions between data and SM predictions~\cite{Alguero:2023jeh,Alguero:2018nvb,Gubernari:2020eft,Gubernari:2022hxn,Altmannshofer:2021qrr,Hurth:2020ehu,Wen:2023pfq,SinghChundawat:2022zdf, SinghChundawat:2022ldm}. Recent investigations~\cite{LHCb:2023gpo,LHCb:2023gel,Bordone:2024hui} accounting for the full-$q^2$ range -- $q^2 = (p_\ell + p_{\bar\ell})^2$ the dilepton invariant mass -- and narrow charmonium resonances have shown that these tensions can be well-described by a $\sim -25\%$ shift to the short-distance physics encoded in the Wilson coefficient (WC) $C_9$, corresponding to the effective operator
\be\label{eq:O9}
 \mathcal{O}_9 = (\bar b_L\gamma^\mu s_L)(\bar\ell \gamma_\mu \ell)\,.
\ee

While the discrepancies between SM predictions and data are clear, the source of the disagreement is challenging to pinpoint. In particular, long-distance contamination arising from the rescattering of intermediate $c\bar c$ states can potentially produce identical effects as those from shifts in $C_9$~\cite{Jager:2014rwa,Ciuchini:2022wbq,Ciuchini:2020gvn,Ciuchini:2019usw}. These non-local effects were most recently estimated in Refs.~\cite{Gubernari:2020eft,Gubernari:2022hxn}, following the approach in 
Ref.~\cite{Khodjamirian:2012rm}. The method relies on an operator-product expansion combined with a hadronic dispersion relation evaluated in the unphysical negative-$q^2$ region. The adopted parameterization satisfies unitarity bounds, and the resulting contribution was found to be small. However, the parameterizations used in Refs.~\cite{Gubernari:2020eft,Gubernari:2022hxn} do not account for anomalous branch cuts which extend into the complex-$q^2$ plane, which arise from triangle-loop graphs corresponding to the rescattering of intermediate states containing charmed mesons. Several investigations into formally quantifying the magnitude of these effects have been made, e.g. through novel parameterizations for the extrapolation of unphysical results~\cite{Gopal:2024mgb} or use of dispersive methods~\cite{Mutke:2024tww}. In Ref.~\cite{Mutke:2024tww}, it was found that the anomalous cuts in pion rescattering in $B\to (\pi,\rho)\gamma^*$ decays could give contributions that are $O(10\%)$ of the non-anomalous non-local form factors, but this cannot be trivially extended to the charm case due to the qualitatively different analytical structure. In fact, in the charm-loop case, the anomalous branch point always lies in the lower complex half-plane  (except for the case of an external pion). Since anomalous effects were found to be the smallest in the light-quark loop under similar conditions, the naive expectation is that their impact is small. Still, an explicit calculation is necessary to confirm this.

In Ref.~\cite{Isidori:2024lng}, we took a different approach by performing a direct, albeit model-dependent, estimation of charm rescattering effects in $B^0\to K^0\llpair$, finding that they were not likely to contribute more than a $10\%$ effect. This study, however, assumed a high degree of symmetry and neglected several effects which are not necessarily trivially small. The goal of this letter is to extend the work of Ref.~\cite{Isidori:2024lng} to include contributions which can arise from dipole couplings of the photon to intermediate charmed mesons. We further clarify some general conclusions about the sign- and $q^2$- dependence of the effect, following from the structure of the electromagnetic form factors.

The paper is organized as follows. 
The effective model in terms of mesonic fields is introduced in Section \ref{sec:topologies}. The form factors needed to modify the point-like interactions introduced in Sec.~\ref{sec:topologies} and to extend our result to the whole kinematic range are discussed in Sec.~\ref{sec:formfactors}. In particular, the extraction of the dipole form factor from data is described. In Sec.~\ref{sec:results}, the results of our calculations are shown, and a multiplicity factor is calculated that accounts for additional possible intermediate states in the topologies we considered. We also show the combination of this calculation with the results from the monopole contributions found in \cite{Isidori:2024lng} and its effect on $C_9$, accounting for a more general monopole form factor. Sec.~\ref{sec:conclusions} summarizes our results.

\section{Topologies and effective interactions} \label{sec:topologies}
The dynamics of $D^{(*)}_{(s)}$ mesons close to their mass shell are described by the following effective Lagrangian, determined by the Lorentz transformation properties of the mesons, gauge invariance under QED, $SU(3)$ light-flavor symmetry, and heavy-quark spin symmetry (to equate the masses of all charmed mesons):
\begin{equation}
\begin{split}
        \mathcal{L}_{D,\text{free}} = & - \frac{1}{2}\big(\Phi_{D^*}^{\mu\nu}\big)^\dag\,\Phi_{D^*\,\mu\nu} - \frac{1}{2}\big(\Phi_{D^*_s}^{\mu\nu}\big)^\dag\,\Phi_{D^*_s\,\mu\nu} \\[0.5em]
        & +\big(D_\mu \Phi_D\big)^\dag\,D^\mu\Phi_D + \big(D_\mu \Phi_{D_s}\big)^\dag\,D^\mu\Phi_{D_s} \\[0.5em]
        &+m_D^2\big[\big(\Phi_{D^*}^\mu\big)^\dag\Phi_{D^*\,\mu} + \big(\Phi_{D^*_s}^\mu\big)^\dag\Phi_{D^*_s\,\mu}\big] \\[0.5em]
        & - m_D^2\big[\Phi_D^\dag\,\Phi_D + \Phi_{D_s}^\dag\Phi_{D_s}\big]+ \text{h.c.}\,,
\end{split}
\label{eq:QEDint}
\end{equation} 
where we have defined
\begin{equation}
\begin{split}
    &\Phi_V^{\mu\nu} = D^\mu \Phi_V^\nu - D^\nu \Phi_V^\mu\,,\\[0.5em]
    &D_\mu \Phi = \partial_\mu \Phi + i\,e A_\mu \Phi\,,
\end{split}
\end{equation}
for $V = D^*, D_s^*$ and $e$ denoting the positron charge. 

In terms of heavy-hadron fields $H_v$ with velocity $v$, the dipole interaction reads
\be \label{eq:hhdipole}
\mathcal{L} \ni g_{\text{dip}}\mathrm{Tr} \Big[ \bar{H}_v \sigma_{\mu \nu} F^{\mu \nu} H_v \Big],
\ee
where $F_{\mu \nu}$ is the photon field strength and $g_{\text{dip}}$ is a coupling constant. The heavy-hadron fields are defined by
\be
 H_{v,a} = \frac{1 + \slashed{v}}{2}\big({P_{v,a}^*}^\mu\gamma_\mu + i P_{v,a} \gamma_5\big)
\ee
with ${P_{v,a}^*}^\mu$ and $P_{v,a}$ the vector ($D^*_{(s)}$) and pseudoscalar ($D_{(s)}$) meson fields, respectively. In order to obtain the Lagrangian in terms of the fundamental meson fields, we convert to the standard relativistic normalization, ${P_{v,d(s)}^*}^\mu = \sqrt{m_{D^*_{(s)}}}\Phi^\mu_{D^*_{(s)}}$ and $P_{v,d(s)} = \sqrt{m_{D_{(s)}}}\Phi_{D_{(s)}}$. Expanding Eq.~\eqref{eq:hhdipole}, we find the Lagrangian
\be
\begin{aligned}\label{eq:funddipole}
\mathcal{L}_{\mathrm{dip}} = 2 i g_{\mathrm{dip}} \Bigg[& \frac{i}{m_D} \widetilde{F}_{\mu \nu} \Big(\Phi_D^\dagger \Phi_{D^* \mu\nu} + \mathrm{h.c.} \Big) \\ &+ 2 \Phi^\dagger_{D^* \mu} \Phi_{D^* \nu} F^{\mu \nu}  \Bigg],
\end{aligned}
\ee
where $\widetilde{F}_{\mu\nu} = \frac{1}{2} \varepsilon_{\alpha \beta \mu \nu} F^{\alpha \beta}$ is the dual photon field strength.

The weak $B\to D^{(*)}D^{(*)}_s$ transitions are described via the following effective Lagrangians
\be
\begin{aligned}
    \mathcal{L}_{BDD} &= g_{DD} \Phi_B \Phi^\dagger_{D_s}\Phi_D + \mathrm{h.c.} , \\
    \mathcal{L}_{BDD^*} &= g_{DD^*}\big(\Phi_{D_s^*}^{\mu\dag}\,\Phi_D\partial_\mu \Phi_B + \Phi_{D_s}^\dag \Phi_{D^*}^\mu \partial_\mu\Phi_B\big) \\ &+ \text{h.c.},
    \\
    \mathcal{L}_{BD^*D^*} &= g_{1} \Phi_B \Phi_{D_s^*}^{\dagger \mu} \Phi_{D^* \mu} + \frac{g_2}{2 m_D^2} \Phi_B \Phi_{D^{*}_s}^{\dagger \mu \nu} \Phi_{D^* \mu \nu} \\ &+ \frac{g_3}{2 m_D^2}\Phi_B \widetilde{\Phi}_{D_s^*}^{\dagger \mu \nu}\Phi_{D^* \mu \nu} +\mathrm{h.c.} \,.
    \end{aligned}
    \label{eq:BDlagr}
\ee
where again we used heavy-quark spin symmetry along with Lorentz invariance to determine the form of the interaction terms. As we will discuss below, the only relevant coupling turns out to be $g_{DD^*}$, as the topologies associated with $g_{DD}$ and $g_1$, $g_2$, and $g_3$ yield vanishing amplitudes. The coupling $g_{DD^*}$ can be extracted from data: redefining the coupling (as in \cite{Isidori:2024lng}), so that it is dimensionless, as
\begin{equation}
    g_{DD^*} = \sqrt{2}G_F\,|V_{tb}^*V_{ts}| m_B m_D\,\bar{g}\,,
    \label{eq:gDD_dec}
\end{equation}
where $V_{ij}$ denotes the elements  of the Cabibbo-Kobayashi-Maskawa (CKM) matrix, and using the averages of the $B \to D^* D_s$ and $B \to D D_s^*$ branching fractions, we find 
\begin{equation}
\bar{g}\approx 0.04\,.
\end{equation}

The vertices describing the kaon emission from the charmed mesons are determined using heavy-hadron chiral perturbation theory (HHChPT), giving
\begin{equation}\label{eq:ddk}
\begin{aligned}
    \mathcal{L}_{DK} &= \frac{2ig_\pi m_D}{f_K}\big(\Phi_{D^*}^{\mu\dag}\Phi_{D_s}\partial_\mu\Phi_K^\dag - \Phi_D^\dag\Phi_{D_s^*}^\mu\partial_\mu\Phi_K^\dag\big)
    \\ &- \frac{2 g_\pi}{f_K} \Phi_{D^*}^{\alpha \dagger} D^\nu \Phi_{D^*_s}^{\beta} \varepsilon_{\alpha \beta \mu \nu} D^\mu  \Phi_K + \mathrm{h.c.}\, .
    \end{aligned}
\end{equation}
where $f_K = 155.7(3)$ MeV~\cite{PDG} is the kaon decay constant and $g_\pi \approx 0.5$~\cite{Becirevic:2012pf}. By construction, the HHChPT effective description is only valid near the kinematic endpoint ($q^2_{\text{max}} \approx m_B^2$), a range corresponding to soft-kaon emission. In this region, our estimate of the rescattering amplitude is more reliable. To extrapolate to lower values of $q^2$, we introduce appropriate form factors, as discussed below.

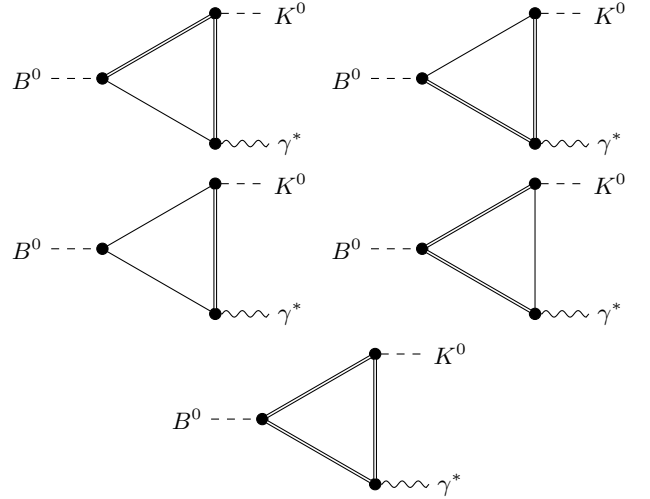
\begin{figure}[t]
        \centering
                \begin{tikzpicture}
                \begin{feynman}
                        \node (i1) {$B^0$};
                        \node[right=1cm of i1, dot] (i2);
                        \node[right=1.5cm of i2] (fake);
                        \node[above=0.866cm of fake, dot] (i3);
                        \node[below=0.866cm of fake, dot] (i4);
                        \node[right=1cm of i3] (j2) {$K^0$};
                        \node[right=1cm of i4] (j3) {$\gamma^*$};
                        \diagram*{
                                (i1) --[scalar] (i2) --[plain] (i4) --[double] (i3) --[double] (i2) ;
                                (i3) --[scalar] (j2) ;
                                (i4) --[photon] (j3) ;
                        };
                \end{feynman}
        \end{tikzpicture}
        \begin{tikzpicture}
                \begin{feynman}
                        \node (i1) {$B^0$};
                        \node[right=1cm of i1, dot] (i2);
                        \node[right=1.5cm of i2] (fake);
                        \node[above=0.866cm of fake, dot] (i3);
                        \node[below=0.866cm of fake, dot] (i4);
                        \node[right=1cm of i3] (j2) {$K^0$};
                        \node[right=1cm of i4] (j3) {$\gamma^*$};
                        \diagram*{
                                (i1) --[scalar] (i2) --[double] (i4) --[double] (i3) --[plain] (i2) ;
                                (i3) --[scalar] (j2) ;
                                (i4) --[photon] (j3) ;
                        };
                \end{feynman}
        \end{tikzpicture}
        \begin{tikzpicture}
                \begin{feynman}
                        \node (i1) {$B^0$};
                        \node[right=1cm of i1, dot] (i2);
                        \node[right=1.5cm of i2] (fake);
                        \node[above=0.866cm of fake, dot] (i3);
                        \node[below=0.866cm of fake, dot] (i4);
                        \node[right=1cm of i3] (j2) {$K^0$};
                        \node[right=1cm of i4] (j3) {$\gamma^*$};
                        \diagram*{
                                (i1) --[scalar] (i2) --[plain] (i4) --[double] (i3) --[plain] (i2) ;
                                (i3) --[scalar] (j2) ;
                                (i4) --[photon] (j3) ;
                        };
                \end{feynman}
        \end{tikzpicture}
                \begin{tikzpicture}
                \begin{feynman}
                        \node (i1) {$B^0$};
                        \node[right=1cm of i1, dot] (i2);
                        \node[right=1.5cm of i2] (fake);
                        \node[above=0.866cm of fake, dot] (i3);
                        \node[below=0.866cm of fake, dot] (i4);
                        \node[right=1cm of i3] (j2) {$K^0$};
                        \node[right=1cm of i4] (j3) {$\gamma^*$};
                        \diagram*{
                                (i1) --[scalar] (i2) --[double] (i4) --[plain] (i3) --[double] (i2) ;
                                (i3) --[scalar] (j2) ;
                                (i4) --[photon] (j3) ;
                        };
                \end{feynman}
        \end{tikzpicture}
                \begin{tikzpicture}
                \begin{feynman}
                        \node (i1) {$B^0$};
                        \node[right=1cm of i1, dot] (i2);
                        \node[right=1.5cm of i2] (fake);
                        \node[above=0.866cm of fake, dot] (i3);
                        \node[below=0.866cm of fake, dot] (i4);
                        \node[right=1cm of i3] (j2) {$K^0$};
                        \node[right=1cm of i4] (j3) {$\gamma^*$};
                        \diagram*{
                                (i1) --[scalar] (i2) --[double] (i4) --[double] (i3) --[double] (i2) ;
                                (i3) --[scalar] (j2) ;
                                (i4) --[photon] (j3) ;
                        };
                \end{feynman}
        \end{tikzpicture}

        \caption{One-loop topologies considered in our analysis. Solid single lines denote charmed pseudoscalars ($D$ or $D_s$) and solid double lines denote charmed vectors ($D^*$ or $D^*_s$). The last three topologies result in vanishing contributions in the fully symmetric limit (see text).}
        \label{fig:loops}
\end{figure}

The topologies relevant to our analysis are shown in Fig.~\ref{fig:loops}.
Those associated with a $DD_s$ intermediate state vanish because of the Lorentz structure of the loop diagram. The topologies associated with a $D^*D^*_s$ intermediate state also vanish when adding together the different diagrams obtained by interchanging a $D^*$ with a $D^*_s$ when using $SU(3)$ chiral symmetry, due to the reversed charge flow in the loop. Therefore, the only non-vanishing topologies turn out to be those associated with $D D^*_s$ and $D_s D^*$ intermediate states, amounting to four diagrams in total to evaluate. In the $SU(3)$-symmetric limit, the two diagrams for each topology give an identical contribution, resulting in a factor of two for each topology.
A discussion on the additional possible intermediate states is presented in Subsec.~\ref{sec:multiplicity}.
The numerical inputs used throughout this paper are the same as in Table I of \cite{Isidori:2024lng}.

\section{Form factors}\label{sec:formfactors}
\subsection{Electromagnetic dipole form factor}
The point-like dipole interactions in Eqs.~\eqref{eq:hhdipole} and~\eqref{eq:funddipole} must be augmented away from the $q^2\to 0$ endpoint to account for the structure of the mesons. As discussed in Ref.~\cite{Isidori:2024lng}, the dominant contribution will arise from the vector charmonium states which can mix with the photon. This can be modeled using a tower of resonances
\be \label{eq:dipoleff}
g_{\mathrm{dip}}(q^2) = \sum_{V} M_V^2 \frac{\eta_V}{q^2-M_V^2 + i \sqrt{q^2} \Gamma_V} e^{i \phi_V} \, ,
\ee
where the sum runs over charmonium resonances.

As $s \to \infty$, $g_{\mathrm{dip}}(s)$ must fall off at least as $s^{-2}$, in order to avoid the violation of unitarity. This condition imposes the following relation:
\begin{equation}
    \sum_{V} M_V^2 \, \eta_V \, e^{i \phi_V} = 0\, , 
\end{equation}
which removes two free parameters (one real and one imaginary). 

The $\eta_V$s and $\phi_V$s can be determined by a fit to data by Belle \cite{Belle:2006hvs} on the $e^+ e^- \to D D^*$ cross section. The scattering $e^+(p_1) e^-(p_2) \to D(p_3) D^*(p_4)$ occurs at tree level only via the $s$ channel. The matrix element is 
\be
\mathcal{M} = \frac{4  e g_{\mathrm{dip}}(s)}{m_D s} \varepsilon_{\mu \nu \alpha \beta}~ p_4^\mu k^\alpha \epsilon^{*}(p_4)^\nu \bar{v}(p_2) \gamma^\beta u(p_1) \, ,
\ee
where $s$ is the center-of-mass energy, yielding a cross section of
\be
\sigma(e^+ e^- \to D D^*) = \frac{e^2 |g_{\mathrm{dip}}(s)|^2}{6 \pi m_D^2} \Big(\frac{s-4 m_D^2}{s}\Big)^{3/2} \, .
\ee
Data on the transition $D^* \to D \gamma$ in the real photon limit can also be used to fit the dipole form factor. Since the two available lattice calculations \cite{Meng:2024gpd, Donald:2013sra} are in agreement with each other, but in disagreement with the only experimental data from CLEO \cite{CLEO:1997rew}, we choose to use an average of the former. The partial width is
\begin{equation}
    \Gamma(D^* \to D \gamma) = \frac{1}{6 \pi m_D^2} |g_{\mathrm{dip}}(0)|^2 \frac{(m_{D^*}^2 - m_D^2)^3}{m_{D^*}^3}  \, .
\end{equation}
The observed value of $\Gamma(D^*\to D \gamma)$ fixes the value of $|g_{\mathrm{dip}}|$ at $q^2=0$, thereby removing one free parameter from the fit. Additionally, Eqs.~\eqref{eq:hhdipole} and~\eqref{eq:funddipole} are valid in the $q^2\to 0$ limit, in which case $\text{Im}g_{\text{dip}}(0) = 0$ to maintain the hermiticity of the effective Hamiltonian, by doing so eliminating an extra free parameter. Limiting the sum over the resonances to $V = J/\Psi, \Psi(2S), \Psi(3770)$, four of the six fit parameters are eliminated by the constraints mentioned above. Furthermore, one overall phase (we fix it to be $\phi_{J/\Psi}$) is irrelevant when fitting the cross-section and can be fixed to $\pi$. Therefore, only one free parameter needs to be determined from the fit. We use the region away from the resonances for the fit ($s>16.5~ \text{GeV}^2$).

It is worth mentioning that the results from the fit are not unique. One set of possible results of the $\chi^2$-fit is shown in Table~\ref{tab:fitdipoleff}. The fit to the data is not optimal, as the reduced chi-squared value turns out to be $\chi^2/\text{dof} = 2.25$, or equivalently, the $p$-value is $3 \times 10^{-3}$. We consider this not to be a problem, since we are mostly interested in describing the behavior of the dipole form factor at high-$q^2$, and are not concerned about precisely modeling the resonance region. These are also partly the reasons why we did not use the results from the fit in Ref.~\cite{Zhang:2010zv}, besides the fact that the authors use the experimental result on $D^* \to D \gamma$ rather than the lattice ones.\footnote{In principle, we could have also extracted the dipole form factor from BESIII data on $B \to D^*_s D^*_s$ \cite{BESIII:2023wsc}. However, these data also receive a monopole contribution, making it difficult to disentangle the two effects.}

\begin{table}[t]
    \centering
    \renewcommand{\arraystretch}{1.25}
    \begin{tabular}{|c|c|c|c|}
    \hline
    $ $ & $J/\Psi$ & $\Psi(2S)$ & $\Psi(3770)$ \\
    \hline
    $ \eta_V$ & $0.3$  & $1.3$ & $1.4$ \\
    $\phi_V$ & $\pi$ & $\pi$ & $0$ \\
    \hline
    \end{tabular}
    \caption{Fit results for the parameters characterizing the electromagnetic dipole form factor.}
    \label{tab:fitdipoleff}
\end{table}

\subsection{Form factor for the \texorpdfstring{$D D^* \to \llpair K$ amplitude}{}}
As mentioned before, our estimate is most reliable in the high-$q^2$ region, as it is based on heavy-hadron chiral perturbation theory, valid in the limit of soft pseudo-Nambu Goldstone meson emission. As in \cite{Isidori:2024lng}, in order to extrapolate our result to the whole kinematic range, we introduce a form factor that modifies the $D D^* \to \llpair K$ amplitude. The HHChPT Lagrangian in Eq.~(\ref{eq:ddk}) leads to an amplitude that grows as $E_K/f_K$ and that is proportional to $1/f_K$, features that are accurate only in the soft-kaon limit. This behavior can be corrected by the following replacement:
\begin{align}
\label{eq:kaonFF}
    \frac{1}{f_K} &\to \frac{1}{f_K} G_K(q^2)\,,
    \\
    G_K(q^2) &= \frac{1}{1 + E_K(q^2)/f_K}
    = \frac{2 m_B f_K}{2m_B f_K + m_B^2 - q^2}\,.
    \nonumber
\end{align}
Although the form factor in Eq.~\eqref{eq:kaonFF} is based primarily on scaling arguments, it turns out to behave very similarly to the local vector form factor, $f_+(q^2)$. This can result in potentially flat, ``$C_9$-like'' contributions in our result for the non-local matrix element, as hypothesized in Refs.~\cite{Jager:2014rwa,Ciuchini:2019usw,Ciuchini:2020gvn,Ciuchini:2022wbq}, particularly away from the resonance region where the photon form factor has less impact. See Ref.~\cite{Isidori:2024lng} for further details.

\section{Results}\label{sec:results}
\subsection{Dipole amplitude}
All the ingredients necessary to estimate the first and second diagrams in Fig.~\ref{fig:loops} are now in place. We use the effective Lagrangians introduced in Eqs.~(\ref{eq:QEDint}), (\ref{eq:funddipole}), (\ref{eq:BDlagr}), (\ref{eq:ddk}), and the dipole form factor in Eq.~(\ref{eq:dipoleff}) with the $\eta_V$s and $\phi_V$s obtained from the fit to the $\sigma(e^+ e^- \to D D^*)$ and $\Gamma(D^* \to D \gamma)$ data.
We will compare the rescattering effects with the short-distance contribution generated by the operator $\mathcal{O}_9$, which has the exact same Lorentz structure and reads
\begin{equation}
\label{eq:noLDCharm}
    \mathcal{M}_{\rm SD} = \frac{4 G_F}{\sqrt{2}}\frac{e}{16\pi^2}V^*_{tb} V_{ts} (p_B\cdot j_{\text{em}})
    f_+(q^2)
    (2 C_9 ), 
\end{equation}
where $f_+(q^2)$ is the $B\to K$ 
vector form factor~\cite{Parrott:2022rgu}.

The sum of the non-vanishing one-loop diagrams in Fig.~\ref{fig:loops} shows an ultraviolet divergence, which is discarded via an $\overline{\text{MS}}$-like renormalization scheme. The resulting renormalization scale- and scheme-dependence are treated as an uncertainty band on the dispersive part of the final result.

The absorptive part of the amplitude is independent of the renormalization scheme used, and it corresponds to the analytical discontinuity that appears in the kinematic region where the internal mesons go on-shell. As expected by gauge invariance, our result is regular for $q^2 \to 0$.

\begin{figure} [h]
    \centering
    \includegraphics[width=0.97\linewidth]{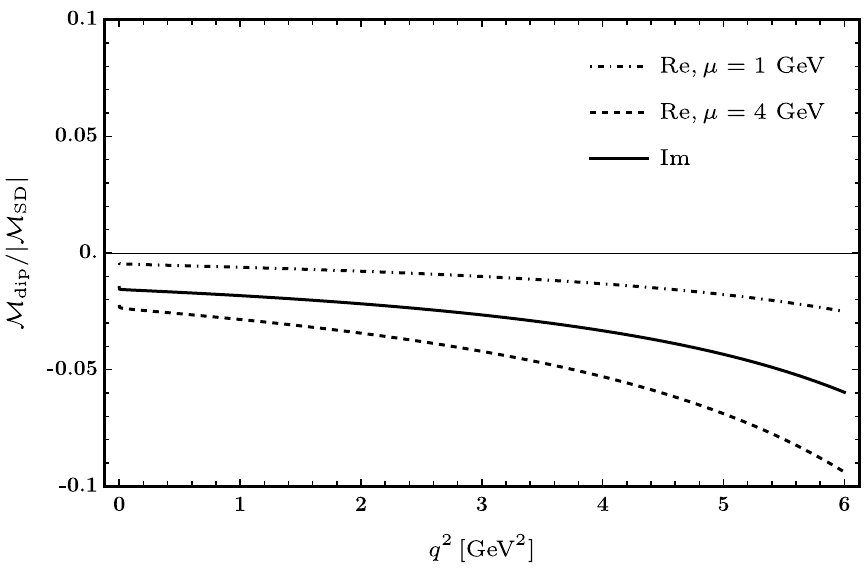}
    \includegraphics[width=0.97\linewidth]{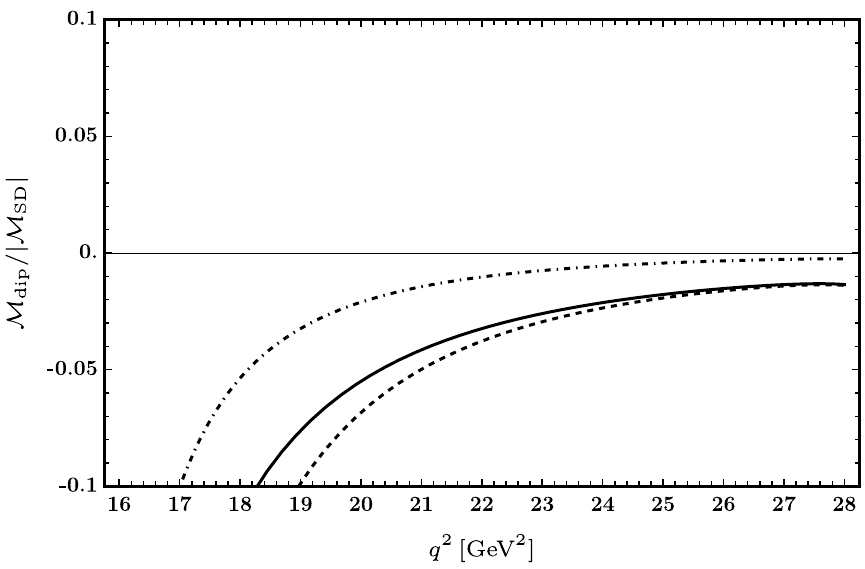}
    \caption{Ratio of the real and imaginary parts of the dipole matrix element over the absolute value of the short-distance matrix element in the low-$q^2$ (top) and high-$q^2$ (bottom) regions. The dashed-dotted and the dashed lines correspond to values of $\mu=1, 4$ GeV for the renormalization scale, respectively.}
    \label{fig:ratioMdipMSD}
\end{figure}

The ratio of the matrix element associated with the rescattering effects due to the dipole interactions over the short-distance matrix element of $B \to K \llpair$ is plotted in Fig.~\ref{fig:ratioMdipMSD}, in the low- and high-$q^2$ (top and bottom plot). The solid line represents the scale-independent absorptive part of the matrix element (over the absolute value of the short-distance matrix element), while the dashed-dotted and the dotted lines represent the dispersive part for two values of the renormalization scale, 1 and 4 GeV. 

The analytical result is given in the Appendix \ref{appendix}.

\subsection{Multiplicity factor for the dipole amplitude}\label{sec:multiplicity}
So far, we have focused on the $D D_s^*$ and $D^* D_s$ intermediate states. However, other possible intermediate states with $\bar{c}c\bar{s}d$ valence structure could contribute to the total rescattering effect. In this section, we account for the additional intermediate states to give a more realistic estimate of the rescattering effects.

The possible intermediate states are found in the following way. Let us define $A_1$ the charmed (or charmed-strange) meson state connecting the $B$ meson to the kaon, $A_2$ the one connecting the $B$ meson to the photon, and $X$ the one connecting the photon to the kaon. We list the largest $B \to A_1 A_2$ decays in Table \ref{tab:BDDdecays}\footnote{Other decays with $\{A_1 A_2\}$ having the same $\bar{c}c\bar{s}d$ valence structure are present, e.g. $B \to K J/\Psi$, but have smaller branching ratios than the listed ones, and give a numerically small contribution to the multiplicity factor.}. 
The idea is to consider the lowest resonant state $X$ that is allowed for any set of $A_1, A_2$ states, since the photon interaction is expected to be suppressed for higher resonances.
For every set of states $A_1,A_2$, the criteria that the intermediate states must satisfy are as follows.
First of all, the kaon (strong) interaction must conserve parity. Therefore:
\be
P(A_1) = (-1)^{L+1} P(X) \, ,
\label{eq:parity}
\ee
where $P$ denotes the eigenvalue under parity and $L$ the orbital angular momentum of the final state.
Furthermore, the photon dipole interaction must conserve parity and requires $L=1$ in the final state. This implies:
\be
P(A_2) = (-1)^{L+1} P(X) = P(X).
\ee
From this last relation, one immediately finds the intrinsic parity of the $X$ state. By using Eq.~(\ref{eq:parity}), one then finds if $L$ must be even or odd. Now, imposing that the total angular momentum must be conserved at the kaon vertex, one can find the allowed values of the spin of the state $X$. Now one can identify the state $X$ among the possible in Table \ref{tab:states}\footnote{
It is worth noting that the $0^+$ and $1^+$ states listed in Table~\ref{tab:states} are very broad resonances as compared to the $D$ and $D^*$. As such, it is unlikely that these unstable states would realistically give substantial contributions to long-distance rescattering effects. However, in an effort to remain conservative, we treat these resonances on equal footing with the $D$ and $D^*$ to compute the multiplicity factor, Eq.~\eqref{eq:multiplicity}.
}. Note that one does not need to verify the conservation of total angular momentum at the photon vertex, as this can be satisfied for any value of $J$ of the state $A_2$.

\begin{table}[t]
    \centering
        \renewcommand{\arraystretch}{1.2}
    \begin{tabular}{|c|c|}
    \hline
        $B^0$ Decay & $\mathcal{B}(B^0\to X)\times 10^3$  \\
         \hline
         $ D D $ &  $7.2\pm 0.8$ \\
         $D^{*}D_s$ & $8.0\pm 1.1$\\
         $D D_s^*$ & $7.4\pm 1.6$\\
         $D^*D_s^*$ & $17.7\pm1.4$\\
         $ D D_{s1}(2460)  $ & $3.5\pm 1.1$\\
         $D^* D_{s1}(2460)$ &  $9.3\pm2.2$\\
         $D D_{s0}^*(2317) $ & $1.06\pm 0.16$ \\
         $D^* D_{s0}^*(2317)$ & $1.5 \pm 0.6 $\\
         \hline
    \end{tabular}
    \caption{List of additional charm-strange $B^0$ decay modes included in the multiplicity factor. All values come from Ref.~\cite{PDG}, and the $D_{sJ}(2460)$ is assumed to be $J^P=1^+$ as indicated in Ref.~\cite{Belle:2003guh}.}
    \label{tab:BDDdecays}
\end{table}

\begin{table}[t]
    \centering
    \renewcommand{\arraystretch}{1.25}
    \begin{tabular}{|c|c|}
    \hline
        $J^P$ & meson state  \\
         \hline
  $0^-$ & $D_{(s)}$ \\
  $1^-$ & $D^*_{(s)}$ \\
  $0^+$ & $D^*_0(2300) / D_{s0}^*(2317)$\\
  $1^+$ & $D_1(2430)$ / $D_{s1}(2460)$ \\
         \hline
    \end{tabular}
    \caption{Lowest resonant charmed or charmed-strange mesons (from Ref.~\cite{PDG}).}
    \label{tab:states}
\end{table}
The first four decays in Table \ref{tab:BDDdecays} correspond to the one-loop topologies that have been explicitly evaluated above. We follow the described procedure for the other decays listed in Table \ref{tab:BDDdecays}. We find that, for the $ D D_{s1}(2460)  $ intermediate states, the possible $X$ states are $D_{s0}^*(2317), D^*$; for the $D^* D_{s1}(2460)$ intermediate states $D_{s1}(2460), D^*$; for the $D D_{s0}^*(2317) $ intermediate states $D_{s0}^*(2317), D_0^*(2300)$; for the $D^* D_{s0}^*(2317)$ intermediate states $D, D_{s1}(2460)$.
In order to estimate the impact of these additional discontinuities, we normalize the $B^0\to X_{\bar{c}c\bar{s}d}$ rates to the $B^0\to D^*D_s+
DD^*_s$ one, assuming that each set of gauge-invariant diagrams with a set of additional intermediate states roughly scales with the size of the corresponding $B^0\to X_{\bar{c}c\bar{s}d}$ amplitude. The multiplicity factor $\mathcal{N}$ is then defined as:
\begin{equation}\label{eq:multiplicity}
    \begin{split}
    \mathcal{N} &= \frac{\sum_{\{A_1 A_2\}}\mathcal{M}(B^0\to A_1 A_2)}{\mathcal{M}(B^0\to D^*D_s) + \mathcal{M}(B^0\to DD^*_s)} \\[0.5em]
    &\approx \frac{1}{2}\sum_{\{A_1 A_2\}}\sqrt{\frac{\mathcal{B}(B^0\to A_1 A_2)}{\mathcal{B}(B^0\to DD_s^*)}} \approx 2.2 \, .
    \end{split}
\end{equation}

\subsection{Reanalysis of the electromagnetic\\ vector form factor}

Our final goal is to combine the results of 
of this work with those presented Ref.~\cite{Isidori:2024lng}, where only the electromagnetic vector form factor (monopole term) has been considered.
To this end, we improve upon Ref.~\cite{Isidori:2024lng} as far as the parameterization of the 
electromagnetic form factor is concerned. More precisely, we adopt a functional form similar to that in Eq.~\eqref{eq:dipoleff}, replacing the electric charge in the covariant derivatives via $e\to e F_V(q^2)$, where
\begin{equation}\label{eq:vecFF}
    F_V(q^2)= \sum_V M_V^2\frac{y_V}{q^2 - M_V^2 + i\sqrt{q^2}\Gamma_V}e^{i\varphi_V}\,,
\end{equation}
as opposed to the na\"ive vector meson dominance (VMD) ansatz used in Ref.~\cite{Isidori:2024lng}. The vector form factor in Eq.~\eqref{eq:vecFF} was then fit using BESIII data for $e^+ e^-\to D_s^+ D_s^-$~\cite{BESIII:2024zdh} subject to the constraint
\begin{equation}\label{eq:lowq2vecFFcontraint}
    F_V(q^2 \to 0) \to 1\, \Rightarrow\, \sum_V y_Ve^{i\varphi_V} = -1\,.
\end{equation}
Note that, unlike $g_{\text{dip}}$, $F_V$ does not have any constraints from unitarity, since the cross-section
\begin{equation}
    \sigma(e^+e^-\to D^+D^-) = \frac{\pi\alpha^2}{3 s}|F_V(s)|^2\Bigg(\frac{s - 4 m_D^2}{s}\Bigg)^{3/2}\,,
\end{equation}
is manifestly unitary as $s\to\infty$. When including $V\in\{J/\psi,\psi(2S),\psi(3770)\}$, this leaves two independent parameters that we fit to the BESIII data away from the resonance region, i.e. for $s>16.5$ GeV$^2$, as in the dipole form factor case. The fit results are shown in Table \ref{tab:fitmonopoleff}. The fit is good enough for our purposes (the reduced $\chi^2$ is $\sim 1.3$ and the p-value is $\sim 0.045$).

\begin{table}[t]
    \centering
    \renewcommand{\arraystretch}{1.25}
    \begin{tabular}{|c|c|c|c|}
    \hline
    $ $ & $J/\Psi$ & $\Psi(2S)$ & $\Psi(3770)$ \\
    \hline
    $ y_V$ & $1.50$  & $1.01$ & $0.63$ \\
    $\varphi_V$ & $\pi$ & $5.75$ & $2.19$ \\
    \hline
    \end{tabular}
    \caption{Fit results for the parameters characterizing the electromagnetic monopole form factor.}
    \label{tab:fitmonopoleff}
\end{table}

It is worth mentioning that also data on $e^+ e^- \to  D^+ D^-$ is available from Belle \cite{Belle:2007qxm}, but with larger uncertainties. For this reason, we chose to use the BESIII data.

In Ref.~\cite{Isidori:2024lng}, it was argued that the vector form factor specifically leads to qualitatively different behavior between the low- and high-$q^2$ regions due to a change of sign in the form factor when using the VMD ansatz. Thanks to the most general structure of the form factor presented here, we can clarify that this feature is not a result of VMD, but instead is a generic consequence of imposing the constraint in Eq.~\eqref{eq:lowq2vecFFcontraint} on a tower of resonances described by Eq.~\eqref{eq:vecFF}, where the most dominant resonances are localized around some reference mass, $\overline M$ such that
\begin{equation}
    \epsilon_V = \frac{M_V^2}{\overline M^2} - 1 \ll 1\,.
\end{equation}
Expanding to linear order in $\epsilon_V$ and using Eq.~\eqref{eq:lowq2vecFFcontraint}, Eq.~\eqref{eq:vecFF} becomes\footnote{We have additionally assumed that $\Gamma_V\ll \overline M$ and use the fact that we are only interested in the regions away from the resonances to neglect the finite-width effects.}
\begin{equation}\begin{split}\label{eq:expandedVecFF}
   & F_V(q^2) = \frac{\overline M^2}{q^2 - \overline M^2}\times \\[0.5em]
        &\quad\times\Big\{-1 + \frac{q^2}{q^2-\overline M^2}\sum_V y_V e^{i\varphi_V}\epsilon_V + O(\epsilon^2)\Big\}\,.
\end{split}\end{equation}
In the low-$q^2$ region, where $q^2\sim \overline M^2/2$, Eq.~\eqref{eq:expandedVecFF} gives
\begin{equation}\label{eq:vecFFLowLim}
    F_V(q^2_{\text{low}}) \sim 2 + 2\sum_V y_V e^{i\varphi_V}\epsilon_V + O(\epsilon^2)\,,
\end{equation}
while in the high-$q^2$ region with $q^2\sim 2\overline M^2$, this becomes
\begin{equation}\label{eq:vecFFHighLim}
    F_V(q^2_{\text{high}})\sim -1 + 2\sum_V y_V e^{i\varphi_V}\epsilon_V + O(\epsilon^2)\,.
\end{equation}
Taking $\overline M$ to be the average of the masses of the three contributing resonances, $|\epsilon_V|\lesssim O(0.1)$. Comparing Eqs.~\eqref{eq:vecFFLowLim} and~\eqref{eq:vecFFHighLim} then shows that, for $O(1)$ values of $y_V$ and general phases, $\varphi_V$, one can generically expect a large phase difference between the low- and high-$q^2$ regions from the monopole contribution. Of course, this is relaxed in the case of large values of $y_V$ and phases near $\varphi_V\approx 0,\pi$, but this scenario leads to additional fine-tuning to achieve the $q^2\to 0$ requirement in Eq.~\eqref{eq:lowq2vecFFcontraint}. Additionally, even if one accepts the tuning necessary to not give a relative phase difference between the regions, since the same $O(\epsilon)$ factor appears in Eqs.~\eqref{eq:vecFFLowLim} and~\eqref{eq:vecFFHighLim}, it is clear that the low-$q^2$ contribution must be enhanced relative to the high-$q^2$ due to the different $O(\epsilon^0)$ contributions. 

These considerations illustrate well the fact that the two $q^2$ regions provide highly complementary information on long-distance contributions to $B\to K\llpair$. On the numerical side, at the high-$q^2$ endpoint our fit yields
\begin{equation}
    \text{Arg}\big[F_V(m_B^2)\big] = 3.01\,,
\end{equation}
whereas the phase of the form factor is near zero over the whole low-$q^2$ region.

Note that this same analysis does not hold for the dipole form factor, since the low-$q^2$ condition from the $D^*\to D\gamma$ decay corresponds to $|g_{\text{dip}}(0)|\approx 0.02$, which, by a similar argument, gives $g_{\text{dip}}(q^2_{\text{low}})\sim g_{\text{dip}}(q^2_{\text{high}})$.

\begin{figure*}[t]
    \centering
    \includegraphics[width=0.8\linewidth]{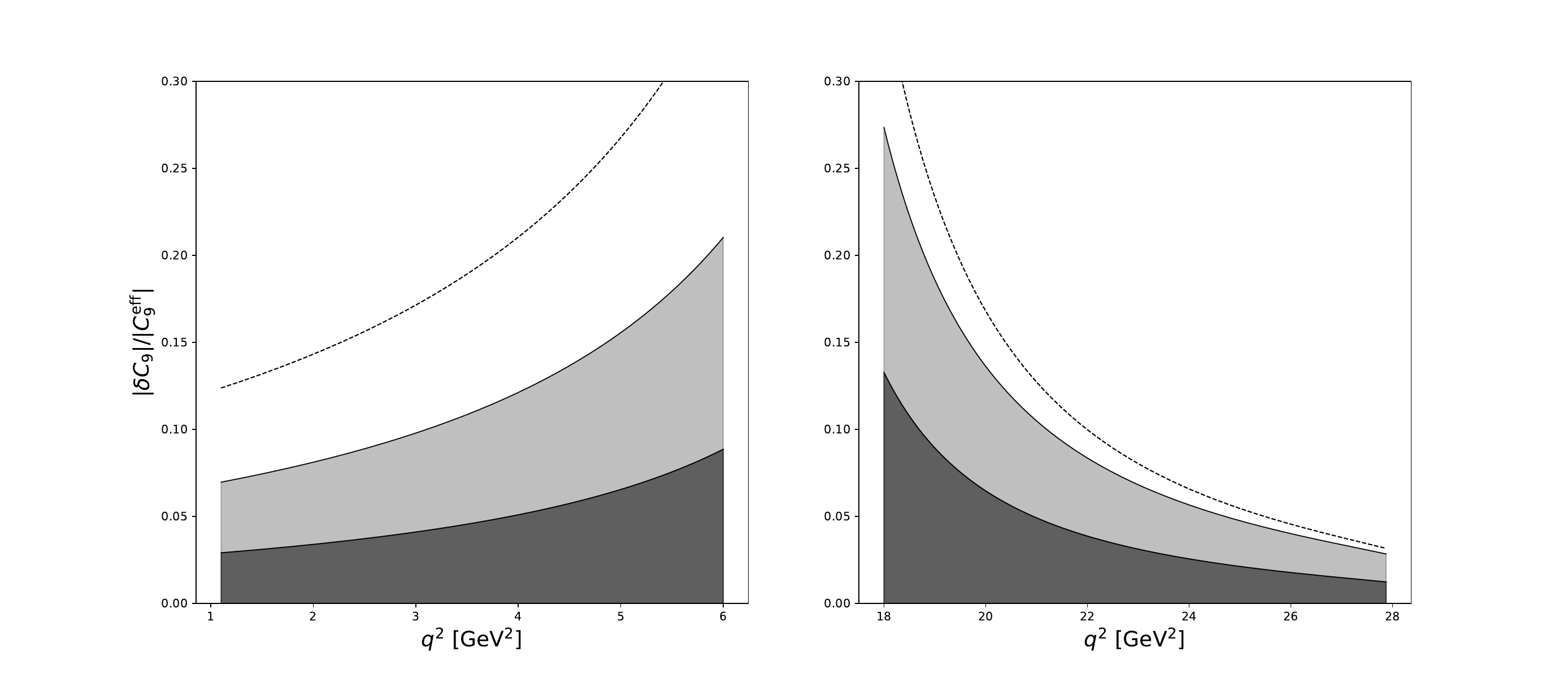}
    \caption{Combined results of this work and that of Ref.~\cite{Isidori:2024lng} plotted as $|\delta C_9|/|C_9^{\text{eff}}|$, where $\delta C_9$ is the contribution from charm rescattering triangle diagrams. The dark gray bands give the ``natural'' results, light gray bands give the partially tuned results, i.e. alignment of all considered $c\bar c d\bar s$ states, and the dashed lines give the fully tuned results, where we assume maximal alignment between all considered intermediate states as well as maximal interference between short-distance and rescattering effects. See text for further details.}
    \label{fig:combinedResults}
\end{figure*}

\subsection{Combined Results}

We are now ready to combine all contributions which interfere with the short-distance SM contribution in the total rate. These can be expressed as an effective shift in the Wilson coefficient $C_9$ that we denote as $\delta C_9$.

Due to the vanishing of diagrams featuring $g_{DD}$, $g_1$, $g_2$, and $g_3$ couplings, all contributions from both dipole and vector interactions enter with the same, single unknown phase of $g_{DD^*}$. Examining Fig.~\ref{fig:ratioMdipMSD} and Fig.~3 of~\cite{Isidori:2024lng}, we find that the absorptive and dispersive pieces of the triangle diagrams are of comparable size. For a generic, untuned phase of $g_{DD^*}$, we therefore expect the contribution to the rate from $DD^*$ rescattering to $\delta C_9$ to be well-estimated by the absorptive piece only. Furthermore, the multiplicity factors introduced in Sec.~\ref{sec:multiplicity} and similarly in Ref.~\cite{Isidori:2024lng} assume that all contributing processes considered are perfectly in phase in order to obtain a conservative upper bound, again requiring a high degree of tuning. We therefore consider three different possibilities for the contributions to $\delta C_9$:
\begin{enumerate}[label = (\alph*)]
    \item \textbf{Natural:} We assume no tuning, in which case $\delta C_9$ is given just by the absorptive part of the $DD^*$ calculations.
    \item \textbf{Multiplicity-Tuned:} We assume no tuning in the phase of $g_{DD^*}$, but tune the relative phases of different intermediate states to exactly add coherently.
    \item \textbf{Fully Tuned:} Same as (b), but also assuming that the phase of $g_{DD^*}$ is aligned for maximal interference with the short-distance $C_9$. Furthermore, we take a relatively large cutoff scale $\mu/m_D \approx 2$.
\end{enumerate}
In all three cases, we do not include effects from the phase differences between low- and high-$q^2$ or relative phases between the monopole and dipole form factors, instead opting to examine the ``worst-case scenarios'', where the monopole and dipole contributions considered in each case always add coherently. As discussed previously, we would not expect this to be the case in general, and in fact, could lead to sizable differences between the low- and high-$q^2$ behavior.

These three possibilities are plotted in Fig.~\ref{fig:combinedResults}. The results show that it is not unfeasible for rescattering effects to give a sizable, $\sim 20\%$ contribution over a large region of $q^2$, at the cost of a more pronounced $q^2$-dependence, contrary to the hypothesis of Refs.~\cite{Ciuchini:2019usw,Ciuchini:2020gvn,Ciuchini:2022wbq} that unaccounted-for non-local contributions to the matrix elements are mimicking short-distance effects, and in tension with the findings of Refs.~\cite{Bordone:2024hui,LHCb:2023gel,LHCb:2023gpo,Alguero:2023jeh} that data are consistent with a $q^2$-independent shift of $C_9$. On the other hand, the ``natural'' case features by far the least $q^2$ variation, and in the regions where the modeling of the resonances has the least impact does not exceed $\sim5\%$.

Of particular interest is the region near the high-$q^2$ endpoint, where the computation suffers least from model-dependent uncertainties. Here, even in the fully tuned case, the contribution to the matrix element again does not exceed $\sim5\%$.

\section{Conclusions}\label{sec:conclusions}

In this work, we have further investigated the role of long-distance rescattering in $B^0\to K^0\llpair$, expanding upon the analysis of Ref.~\cite{Isidori:2024lng} by incorporating effects from dipole interactions of the photon with the rescattered charmed mesons. We find that the dipole contributions from individual rescattered states are comparable in magnitude to the corresponding monopole contributions computed in~\cite{Isidori:2024lng}, but feature a significant dependence on the charmonium resonances in the dipole form factor, even well into the traditional high-$q^2$ region.

When combined with previous results, we find that, under tuned conditions among the different couplings involved,  charm rescattering effects reach up to  $\sim 20\%$ of the short-distance amplitude. However, this not only requires a remarkable degree of tuning to accomplish, but also introduces a strong $q^2$-dependence, whereas experimental data are consistent with a $q^2$-independent shift of $C_9$ with the current experimental uncertainties. The most natural 
case we considered has a weaker dependence on $q^2$, but does not exceed $5\%$ over most of the range. Interestingly, in the region near the high-$q^2$ endpoint, where the calculation is most trustworthy, even the case featuring the highest degree of tuning also gives a $\sim5\%$ contribution.

In summary, our findings suggest that a tension arises when attempting to explain the exclusive mode anomalies solely with unaccounted-for contributions from non-local matrix elements: it is challenging to simultaneously generate an effect which is large enough to explain the anomalies while maintaining the short-distance-like structure (i.e.~no channel- and $q^2$- dependence) favored by present data.

Of course, the model-dependence of our results should not be understated. The results presented rely heavily on the choice of form factors, and the method of accounting for additional $c\bar c d \bar s$ intermediate states is quite rough.
However, the fact that non-local matrix elements could be bounded by future detailed  $q^2$-dependent data analyses emerges as a general conclusion.

\subsection*{Acknowledgments}
We would like to thank Nico Gubernari for helpful comments and discussions.
This project has received funding by the Swiss National Science Foundation~(SNF) under contract~200020\_204428.

\appendix
\section{Analytical results}\label{appendix}
Here, we present the analytical results for the long-distance matrix elements calculated using the model described in the previous sections. 
Let us define
\begin{align}
&L_\mu~=~\log(\mu^2/m_D^2) \, , \\[0.5em]
&l(x, y) = \log{\Big( \frac{2 y - x + \sqrt{x(x-4y)}}{2y}\Big)} \,, \\[0.5em]
        &L(x, y) = l(x, y) \Big[\sqrt{x(x - 4 y)} + y ~ l(x, y)\Big]\, , \\[0.5em]
        &\delta L(q^2, m_B^2, m_D^2) = \frac{L(m_B^2, m_D^2) - L(q^2, m_D^2)}{q^2 - m_B^2}\,.
\end{align}

The matrix element of the charm-rescattering associated with the first topology in Fig.~\ref{fig:loops} reads
\begin{equation}\label{eq:result1}
    \begin{aligned}
        \mathcal{M}^{(1)}_{\text{LD}} &= \frac{i g_{DD^*} g_\pi g_{\text{dip}}(q^2) G_K(q^2) }{2 f_K m_D \pi^2 } (p_B\cdot j_{\text{em}})\times
        \\[0.5em]
        &\times\Big[\big(3 + L_\mu\big)
        - \delta L(q^2, m_B^2, m_D^2)\Big] \, .
   \end{aligned}
\end{equation}

The matrix element of the charm-rescattering associated with the second topology in Fig.~\ref{fig:loops} has a more complicated analytical structure:
\begin{widetext}
\begin{equation}\label{eq:result2}
    \begin{split}
        \mathcal{M}^{(2)}_{\text{LD}} &= \frac{i g_{DD^*} g_\pi g_{\text{dip}}(q^2) G_K(q^2) }{2 f_K m_D \pi^2 } (p_B\cdot j_{\text{em}})
        \times \Bigg[ \frac{9 m_B^2 - 12 m_D^2 + 4 q^2}{18 m_D^2} + \frac{3 m_B^2 - 12 m_D^2 + q^2}{12 m_D^2} L_\mu \\ &+\frac{1}{m_B^2-q^2} \Bigg( \frac{m_B^4 + 4 m_D^2 q^2-8 m_B^2 m_D^2 - m_B^2 q^2}{4 m_D^2} \sqrt{m_B^2 - 4 m_D^2}~l(m_B^2, m_D^2) +\frac{1}{2}(2m_D^2 -m_B^2) ~l(m_B^2, m_D^2)^2 \\ &- \frac{8 q^2 m_D^2+q^4 - 20 m_B^2 m_D^2 - m_B^2 q^2}{12 m_D^2}  \sqrt{q^2-4 m_D^2}~l(m_D^2, q^2) -\frac{1}{2}(2 m_D^2 -m_B^2)~ l(m_D^2, q^2)^2 \Bigg) \Bigg] \, .
    \end{split}
\end{equation}
\end{widetext}

\bibliography{references}
\end{document}